\documentclass[12pt]{article}
\usepackage{epsfig}

\newcommand{\beq}{\begin{equation}}
\newcommand{\eeq}{\end{equation}}
\newcommand{\beqa}{\begin{eqnarray}}
\newcommand{\eeqa}{\end{eqnarray}}

\newcommand{\jpsi}{{J\!/\!\psi}}
\newcommand{\threeSone} {{^3\!S_1}}
\newcommand{\oneSzero} {{^1\!S_0}}

\newcommand{\threePj} {{^3\!P_J}}
\newcommand{\threeSoneo}{{^3\!S_1^{(8)}}}
\newcommand{\oneSzeroo}{{^1\!S_0^{(8)}}}
\newcommand{\threePjo}{{^3\!P_J^{(8)}}}
\newcommand{\threePzeroo}{{^3\!P_0^{(8)}}}

\newcommand{\threePtwoo}{{^3\!P_2^{(8)}}}

\renewcommand{\o}{{\cal O}}

\begin{document}

\begin{titlepage}
\begin{flushright}
DESY 97-091\\
hep-ph/9706374
\end{flushright}

\begin{center}
\vbox{
\vspace{4cm}
PHENOMENOLOGY OF QUARKONIA PRODUCTION IN FIXED TARGET EXPERIMENTS AND AT THE
TEVATRON AND HERA COLLIDERS$^\dagger$
}
\vspace{1cm}

{\sc Matteo Cacciari}\\[3pt]
{\sl \small Deutsches Elektronen-Synchrotron DESY, Hamburg, Germany}\\[-2pt]
{\small cacciari@desy.de}\\[5cm]

{\bf Abstract}\\
\end{center}
The phenomenology of heavy quarkonia production in fixed target experiments and
at the Tevatron and HERA colliders is reviewed. The latest theoretical
results are presented and compared with data, with emphasis on the predictions 
of the factorization approach by Bodwin, Braaten and Lepage.

\vfill
\noindent
\rule{6cm}{.2mm}\\

\vspace{-.7cm}
{\small
\noindent
$^\dagger$ To appear, in a slightly shortened form,
in the Proceedings of the XXXII Rencontres de Moriond, ``QCD and High Energy 
Hadronic Interactions'', Les Arcs, March 1997.
}

\end{titlepage}

\section{Introduction}

The production of heavy quarkonia has been subjected to intense study in the
last two or three years, with tens of papers having being
produced on the problem of $\jpsi$'s, $\chi$'s and $\Upsilon$'s production in
$e^+e^-$, $\gamma p$, $p\bar p$, $pN$, $\pi N$ collisions and also $B$ decays.

The reason for such a surge in interest was the appearance of a theoretical
framework, the so called Factorization Approach (FA) by Bodwin, Braaten and
Lepage \cite{bbl}, which seems able to solve  the theoretical problems
that quarkonia production models faced in the past, and also to reconcile
theoretical predictions with experimental data, previously in disagreement up
to factors of fifty in some instances.

In this talk I shall not review the Factorization approach in detail, leaving
this theoretical introduction to other sources (see for instance 
\cite{bfy,braaten,beneke}). I shall also not discuss in detail the Color
Singlet Model (CSM) \cite{CSM} and the Color Evaporation Model (CEM)\cite{CEM}
(the latter has also been recently compared to data and found able to describe 
at least some of them \cite{sv,aegh}).

I shall just recall how the factorization approach
writes the quarkonium state $H$ production cross section in the following form:
\beq
\sigma(ij\to Q\bar Q\to H) = \sum_n \hat\sigma(ij\to Q\bar
Q[n])\langle\o^H(n)\rangle .
\eeq
According to this equation, the cross section for producing the observable
quarkonium state $H$ is factorized into two steps. In the short distance part a
$Q\bar Q$ pair of heavy quarks is produced in the spin/color/angular momentum
state $^{2S+1}L_J^{(c)} \equiv n$ by the scattering of the two light
partons $i$ and $j$. Successively this pair hadronizes into the quarkonium $H$,
and $\langle\o^H(n)\rangle$ is formally a Non Relativistic QCD (NRQCD) matrix
element describing this non perturbative transition. 

An important feature of
this equation is that also $Q\bar Q$ pairs in a color octet state are allowed
to contribute to the production of a color singlet quarkonium $H$: their color
is neutralized via a non perturbative emission of soft gluons. While the
corresponding matrix elements are suppressed by the need of such an emission,
the short distance coefficients can on the other hand be large, perhaps
overcompensating the suppression of the non perturbative term. This explains
why color octet contributions can play a very important role in predicting the
total size of quarkonia production cross sections.

Many theoretical items would be worth discussing about the Factorization
Approach. Actually, they would probably be worth a seminar (or more) by
themselves. As I said, I shall however skip such a detailed discussion and
rather concentrate on some selected phenomenological outcomes of the
theoretical investigations which have been carried on so far. I shall restrict
myself to analyses of experimental data coming from $p\bar p$ collisions at
the Tevatron, $\gamma p$ collisions at HERA and fixed target experiment,
reviewing the results of these investigations. References will be
provided to the theoretical papers, leaving to them the task of properly
citing the experimental ones. 

The aim of the game will be to check whether the non perturbative matrix
elements which can be extracted from these experimental data are mutually
consistent with each other. In other words, we shall check whether the
factorization approach can properly describe all the data with matrix elements
truly universal and independent  from the underlying short distance process,
as they should.

\section{The Tevatron data in $p\bar p$ collisions}

Beginning with the Tevatron data looks appropriate as the explanation of its 
anomalously large $\psi'$ production rate (a factor of fifty above the Color
Singlet Model prediction) was the first phenomenological breakthrough of the
FA. Braaten and Fleming \cite{bf} explained this large rate by assuming it was
due to a color octet $Q\bar Q$ pair originating via perturbative splitting from
a large $p_T$ gluon. While the non perturbative matrix element for such a color
octet pair to produce a $\psi'$ is predicted by NRQCD to be about two order of
magnitudes suppressed with respect to the one for a color singlet, the
production rate of gluons (and hence of such pairs) is however so large that it
can more than compensate for the suppression. Indeed, Braaten and Fleming
could successfully describe the data fitting a value for the matrix element in
good agreement with the theoretically predicted two-orders-of-magnitude
suppression.

This apparent success of the factorization approach on the $\psi'$ anomaly
problem  made immediately clear the potential importance of color octet
mediated channels and stimulated similar research in other reactions: by the
time of this Conference, the Braaten and Fleming's paper has received more than
120 citations.

\begin{figure}[t]
\begin{center}
\begin{minipage}{8cm}
\epsfig{file=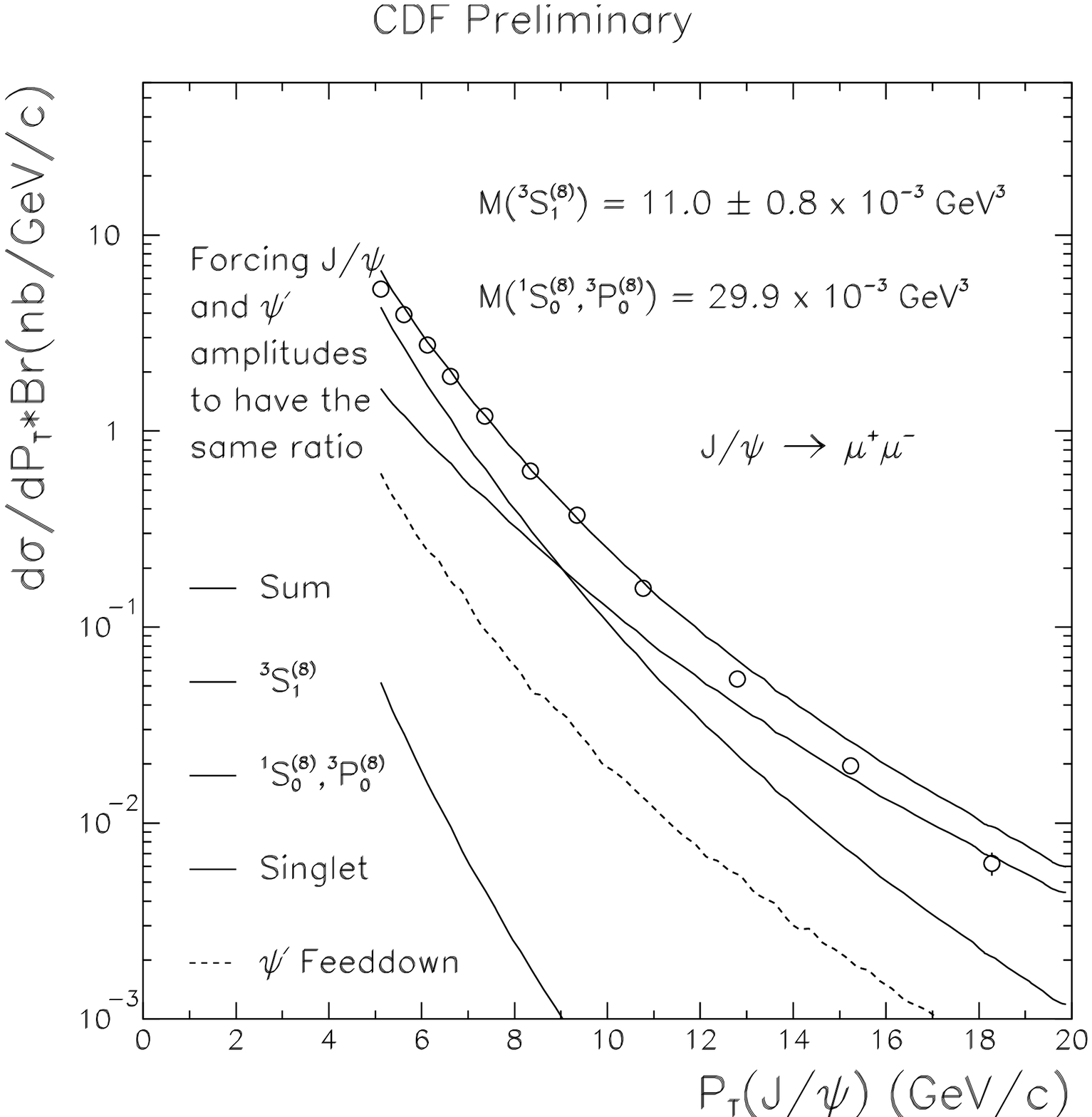,
             height=6cm,width=7cm,
            clip=
}
\end{minipage}
\begin{minipage}{8.cm}
\epsfig{file=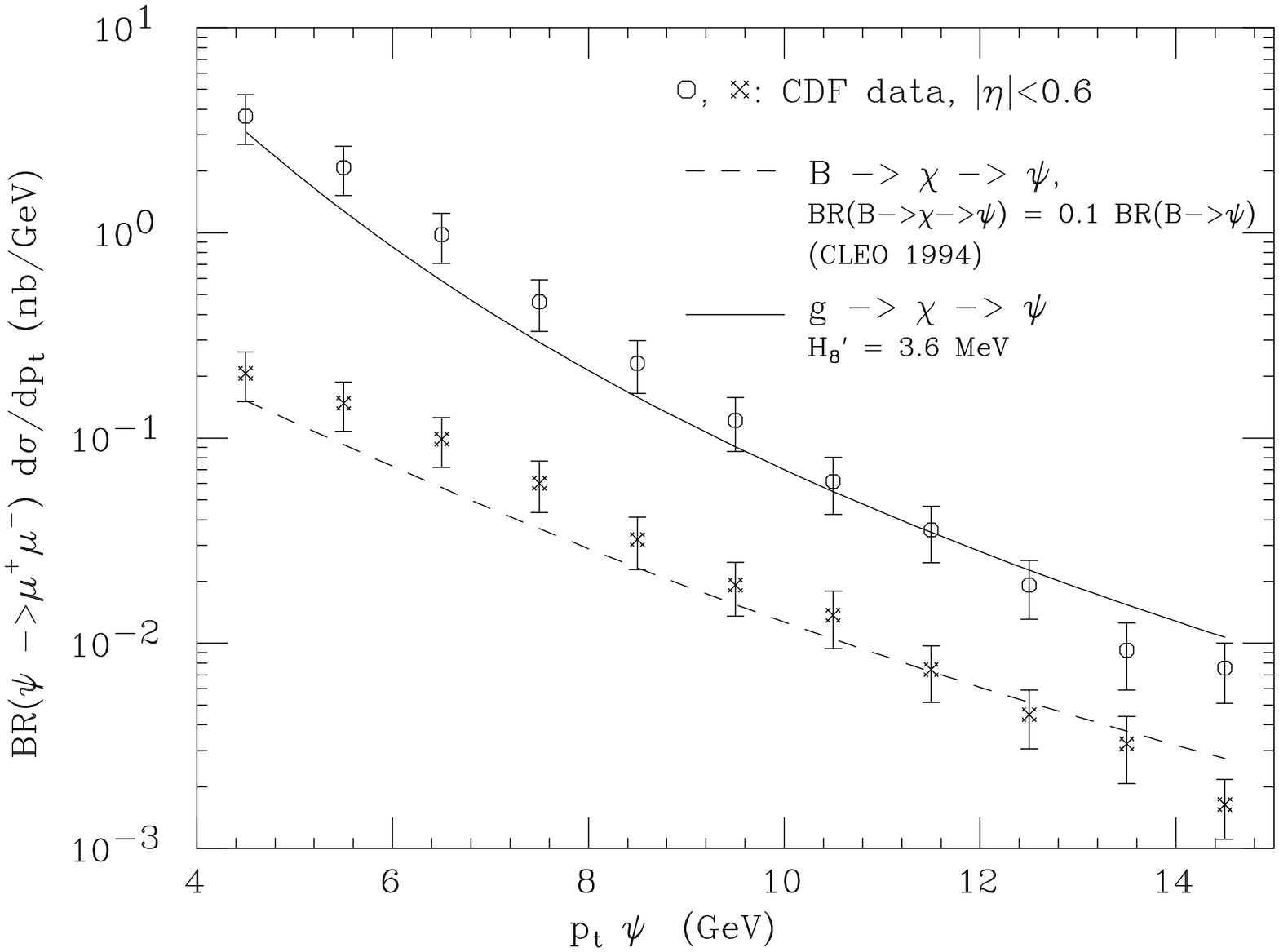,
              bbllx=10pt,bblly=10pt,bburx=600pt,bbury=750pt,
             height=8cm,width=7cm,angle=90,
            clip=
}
\end{minipage}
\begin{minipage}{9cm}
\vspace{20pt}
\epsfig{file=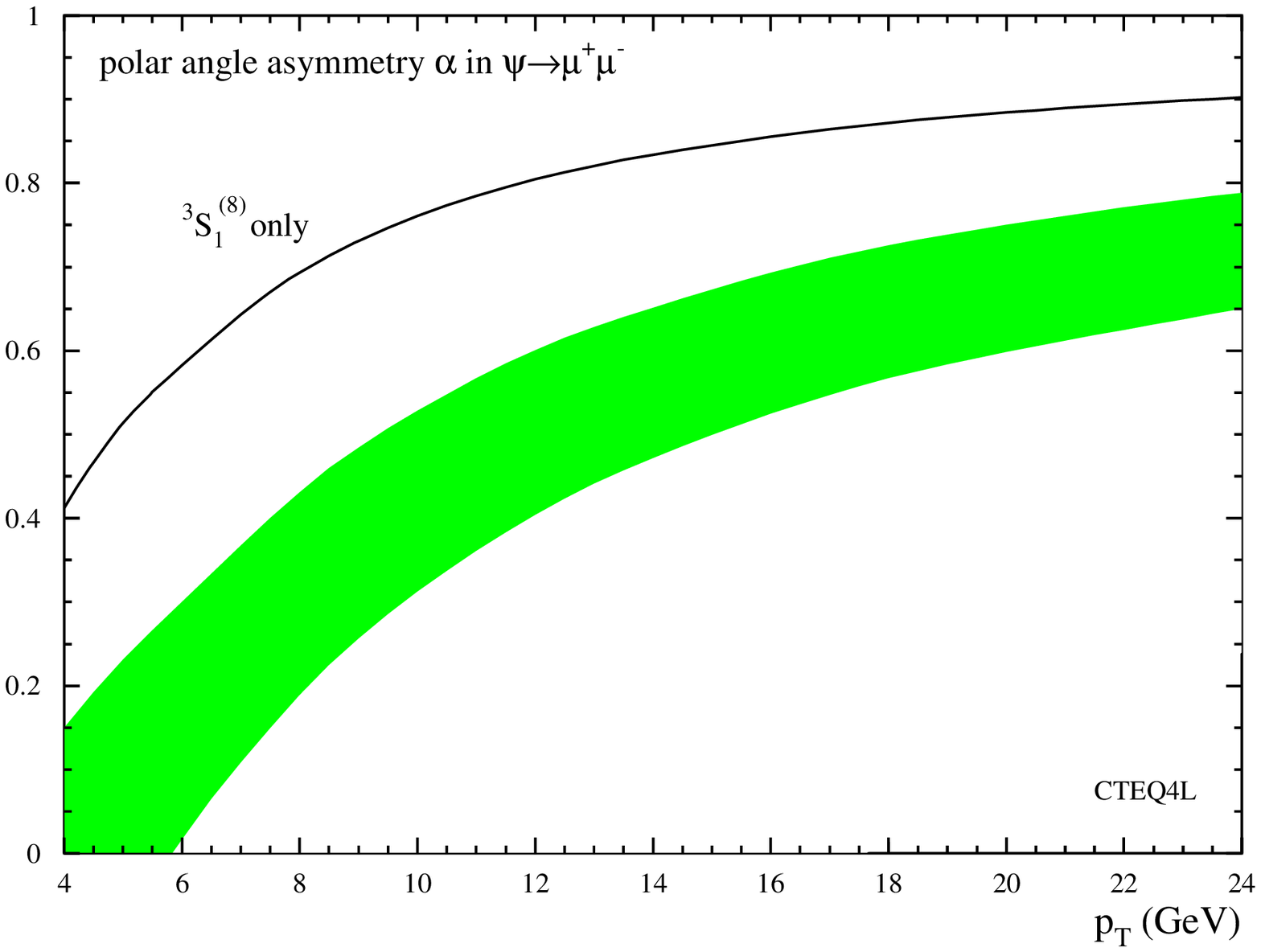,
              bbllx=10pt,bblly=100pt,bburx=540pt,bbury=570pt,
             height=7cm,width=8cm,
            clip=
}
\vspace{-2cm}
\end{minipage}
\caption{\label{fig1}
\small Fits to production of direct $\jpsi$ (left) \protect\cite{cl} 
and $\chi$ (right) \protect\cite{cgmp} at CDF. The $\jpsi$ plot also shows by
how much the Color Singlet Model underestimates the data. Also shown (below) 
is the $\jpsi$ polarization pattern predicted \protect\cite{bk} 
by the factorization approach.}
\end{center}
\end{figure}

Cross sections for large $p_T$ $\jpsi$'s and $\chi$'s production at the
Tevatron have been analyzed, and a more detailed study of $\psi'$ has also been
performed. It was found the theoretical curves could describe the shape of the
data pretty well. The following matrix elements values were returned by the
fits, performed either within the fragmentation approximation 
\cite{PT_TEV,cgmp} or evaluating the full leading order matrix 
elements \cite{cl,bk}:
\beqa
\langle\o^\jpsi_8(\threeSone)\rangle &\simeq& 
(1.5 ~\mbox{\cite{cgmp}}, 1.1 ~\mbox{\cite{st}}, 1.06 ~\mbox{\cite{bk}})
\times 10^{-2}~{\rm GeV}^3\\
\langle\o^\jpsi_8(\oneSzero)\rangle 
+{\textstyle 3\over \textstyle m^2} \langle\o^\jpsi_8(^3P_0)\rangle 
&\simeq& 9 \times 10^{-2}~{\rm GeV}^3
\qquad\mbox{\cite{st}}\\
\langle\o^\jpsi_8(\oneSzero)\rangle 
+{\textstyle 3.5\over \textstyle m^2} \langle\o^\jpsi_8(^3P_0)\rangle 
&\simeq& 4.38 \times 10^{-2}~{\rm GeV}^3
\qquad\mbox{\cite{bk}}\\[20pt]
\langle\o^{\psi'}_8(\threeSone)\rangle &\simeq& 
(4.3 ~\mbox{\cite{cgmp}}, 3.8 ~\mbox{\cite{st}}, 4.4 ~\mbox{\cite{bk}}) 
\times 10^{-3}~{\rm GeV}^3\\
\langle\o^{\psi'}_8(\oneSzero)\rangle 
+{\textstyle 3\over \textstyle m^2} \langle\o^{\psi'}_8(^3P_0)\rangle 
&\simeq& 3 \times 10^{-2}~{\rm GeV}^3
\qquad\mbox{\cite{st}}\\
\langle\o^{\psi'}_8(\oneSzero)\rangle 
+{\textstyle 3.5\over \textstyle m^2} \langle\o^{\psi'}_8(^3P_0)\rangle 
&\simeq& 1.8 \times 10^{-2}~{\rm GeV}^3
\qquad\mbox{\cite{bk}}\\[20pt]
\langle\o^{\chi_J}_8(\threeSone)\rangle &\simeq&
(2J+1) m^2 \times 3.6 \times 10^{-3}~{\rm GeV}^5 
\qquad\mbox{\cite{cgmp}}
\eeqa
$m$ is the charm mass, usually taken equal to 1.5 GeV.
Figure \ref{fig1} shows the CDF data from the Tevatron and the curves which 
fit them with these parameters.

The uncertainties on these fits are certainly not smaller than a factor 
of two, due to
the many systematics entering their determination: parton distribution
functions (responsible for the difference between \cite{st} and \cite{bk}), 
charm quark mass, $\alpha_s$ value, higher order QCD corrections,
etc. They could, however, even be larger. Indications in this direction come
from a fit \cite{ccsl} which makes use of PYTHIA for simulating the effect 
of initial state
radiations from the partons before they collide to produce the $Q\bar Q$ pair:
this changes the slope of the theoretical predictions and hence the result of
the fit:
\beqa
&&\langle\o^\jpsi_8(\threeSone)\rangle \simeq 3 \times 10^{-3}~{\rm
GeV}^3\qquad\mbox{\cite{ccsl}}\\
&&\langle\o^\jpsi_8(\oneSzero)\rangle 
+{3\over m^2} \langle\o^\jpsi_8(^3P_0)\rangle 
\simeq 1.2 \times 10^{-2}~{\rm GeV}^3\qquad\mbox{\cite{ccsl}}
\eeqa
The effect can be seen to be large, the results being significantly
smaller. While such a reduction would actually be welcome in the light of
other data which will be presented further on, one should however for the time
being only take
this as an indication of the size of the uncertainty which may still lay
hidden in their determination.

Other than the size of the production cross section, the polarization of the
quarkonia is an observable with great discriminating power for the various
approaches to quarkonia production. It can be measured by analyzing the
angular distribution of the quarkonium decay products (muons) in its rest
frame, and parametrizing it as
\beq
{d\sigma(\psi\to\mu^+\mu^-)\over{d\cos\theta}} \propto 1 + \alpha(\psi)
\cos^2\theta
\label{pol}
\eeq
$\jpsi$'s produced at the Tevatron at large $p_T$ are predicted to be almost
fully transversely polarized, i.e. $\alpha(\jpsi) \simeq 1$ \cite{br1}, as a
result of the production via gluon fragmentation into $\threeSoneo$ states
being largely dominant. At smaller $p_T$, on the other hand, non-fragmentation
channels involving $\oneSzeroo$ and $\threePjo$ become important: the $\jpsi$'s
are then predicted to be produced essentially unpolarized in the low transverse
momentum region, around $p_T\simeq 5$ GeV \cite{bk}. The observation of such a
polarization pattern, shown in figure \ref{fig1}, would provide great 
support for the factorization approach to quarkonia production.

\section{The HERA data in $\gamma p$ collisions}

Within $\gamma p$ collisions $\jpsi$'s can be produced in leading order at
non-zero $p_T$ via the color singlet channel  $\gamma p \to\, ^3S_1^{(1)}
g\to \jpsi X$ (\cite{CSM}, first reference). Next-to-leading order QCD
corrections to this channel have been recently computed \cite{mk}, and the
results have been found in fairly good agreement with the experimental results 
from
the ZEUS and H1 experiments: the absolute  normalization of the total cross
section agrees within the theoretical uncertainties, and the shape of the 
inelasticity distribution of the $\jpsi$ (usually
denoted by $z$, with $z=E_{\jpsi}/E_\gamma$ in the proton rest frame) is well
described by the calculation.

\begin{figure}
\begin{center}
\begin{minipage}{8cm}
\hspace{-1cm}
\epsfig{file=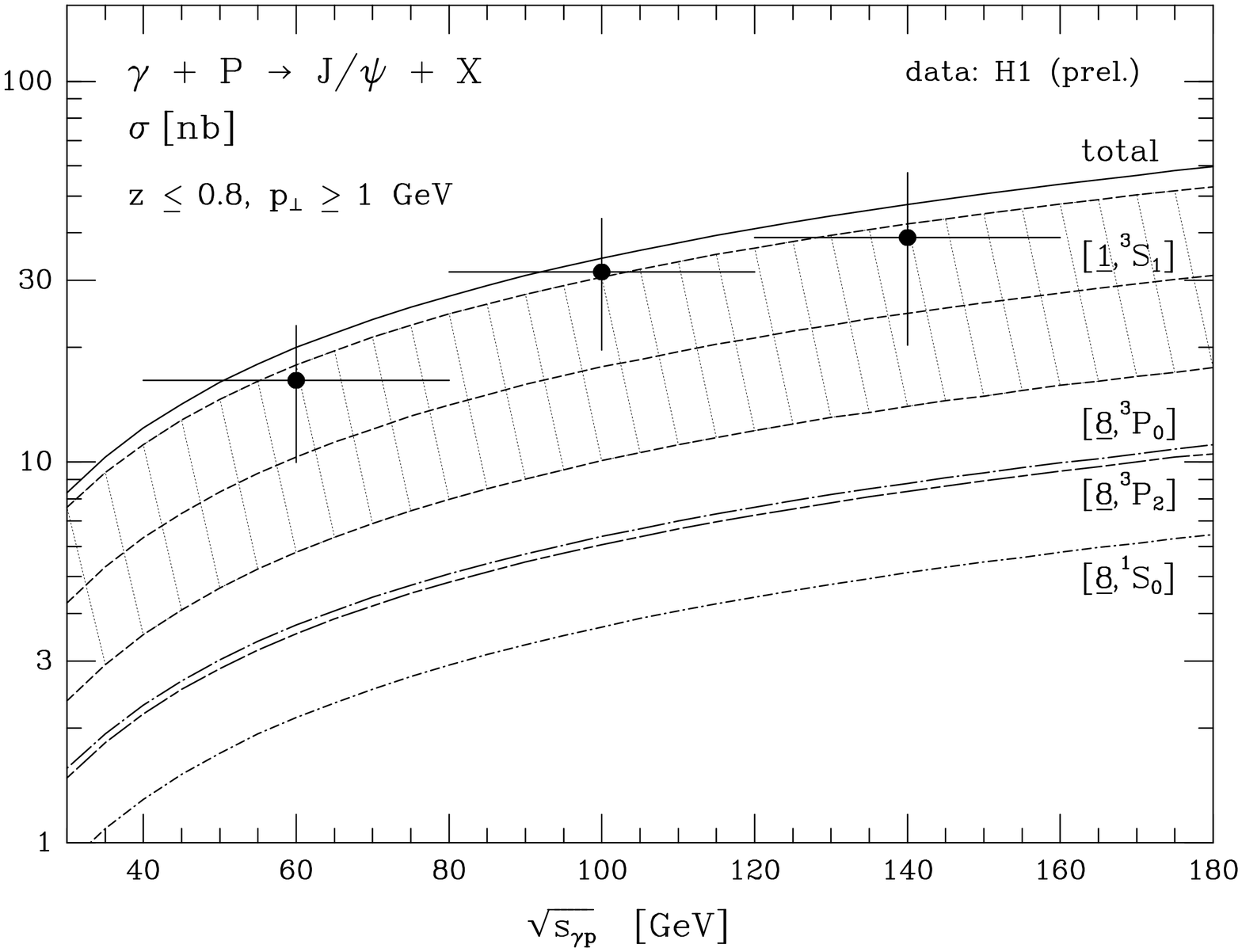,
              bbllx=10pt,bblly=10pt,bburx=550pt,bbury=770pt,
             height=8.5cm,width=7cm,angle=-90,
}
\end{minipage}
\begin{minipage}{8.cm}
\epsfig{file=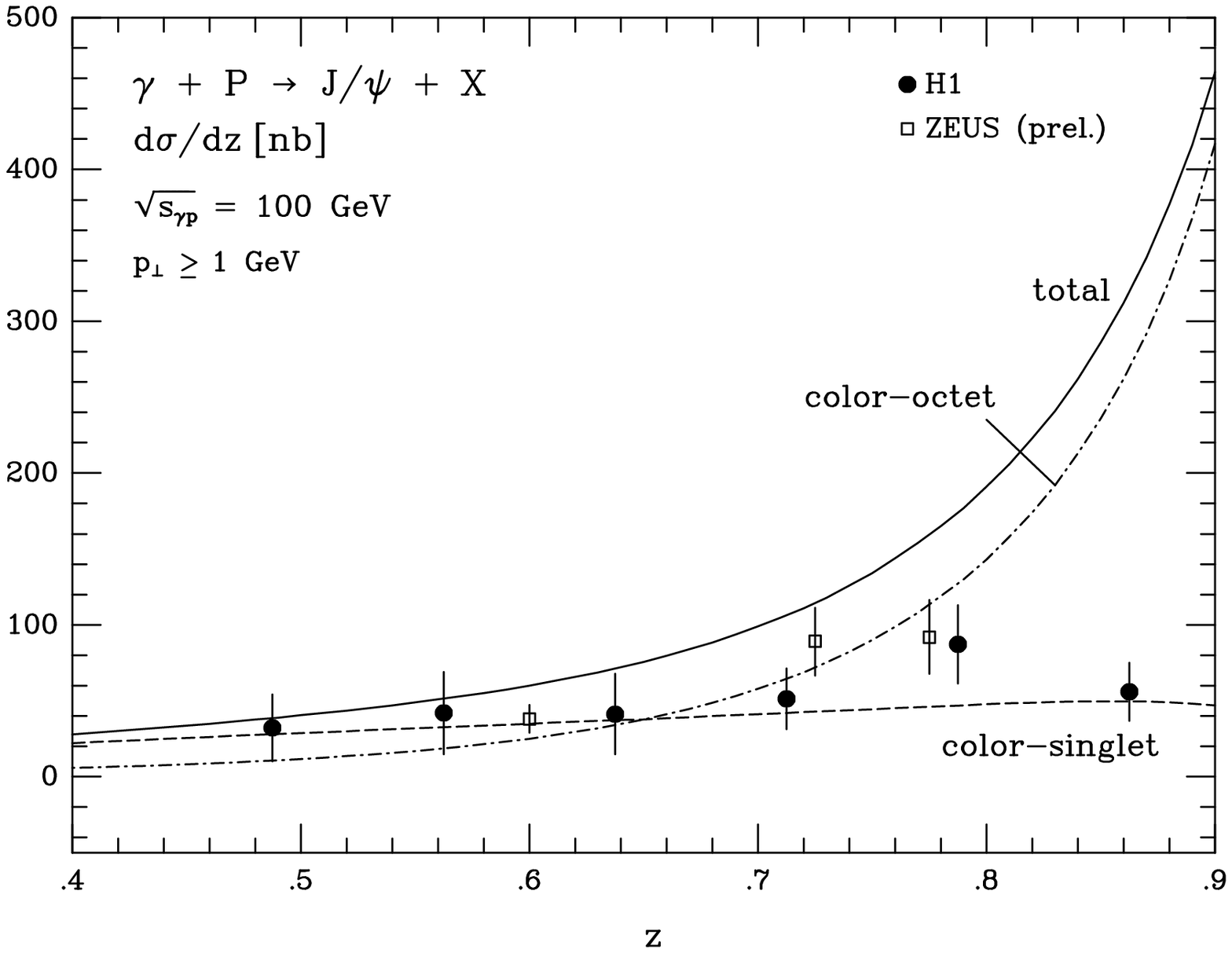,
             height=6.6cm,width=7.7cm,
            clip=
}
\end{minipage}
\caption{\label{fig2}
\small Total cross section (left) and inelasticity distribution
(right) in photoproduction at HERA \protect\cite{ck}. Color octet curves 
with parameters fitted to Tevatron data in \protect\cite{cl}.
}
\end{center}
\end{figure}

Color octet contributions to $\jpsi$ photoproduction have been
investigated in leading order \cite{ck,kls}. In the non-zero $p_T$ region the 
five $\gamma p \to
(\oneSzero^{(8)}, \threeSone^{(8)}, \threePj^{(8)}) g$ channels contribute.
Figure \ref{fig2} shows the results for the total cross section and the
inelasticity distribution with the inclusion of these channels: matrix elements
{\sl of the order} of the ones fitted to the Tevatron data without using PYTHIA
have been employed in these plots, taking $\langle\o^\jpsi_8(\threeSone)\rangle
=  \langle\o^\jpsi_8(\oneSzero)\rangle =
\langle\o^\jpsi_8(\threePj)\rangle/(2J+1)m^2  = 1\times 10^{-2}~{\rm GeV}^3$.

One can see from the plots how the data do not need any octet contributions:
the color singlet channel by itself can describe them well. More than this, the
octet terms evaluated with the Tevatron parameters look at variance with the
data, suggesting a non-universality of these NRQCD matrix elements. We have
however seen how indications exist \cite{ccsl} that the fits to the Tevatron
data may be a significant overestimate: if we reduce the value of the matrix
elements used in the photoproduction calculation by a factor of three,  to
bring them in line with the smaller Tevatron fits returned by using PYTHIA, the
discrepancy in the inelasticity distribution is greately reduced. Further
unaccounted for contributions, like higher orders near the
phase space end point, could easily provide large corrections and bring the
prediction in agreement with the data. See refs. \cite{beneke,brw} for a more
detailed discussion about this point.

An analysis of photoproduction data within the factorization approach has also
been attempted in the elastic region, by fitting experimental results with the
leading order prediction given by the octet channels $\gamma g\to \oneSzeroo,
\threePzeroo, \threePtwoo$. This gives the result \cite{afm}
\beq
\langle\o^\jpsi_8(\oneSzero)\rangle 
+ {7\over m^2} \langle\o^\jpsi_8(^3P_0)\rangle
\simeq 2 \times 10^{-2}~{\rm GeV}^3 \qquad \mbox{\cite{afm}}
\eeq
This looks smaller than some of the Tevatron fits but, as we shall see in  the
next Section, in line with fits to fixed target data. This result should
however be taken with great care, due to the many subtleties surrounding
elastic vector meson production and to the large theoretical uncertainties.

\section{Quarkonia in fixed target experiments}

\begin{figure}
\begin{center}
\begin{minipage}{8cm}
\epsfig{file=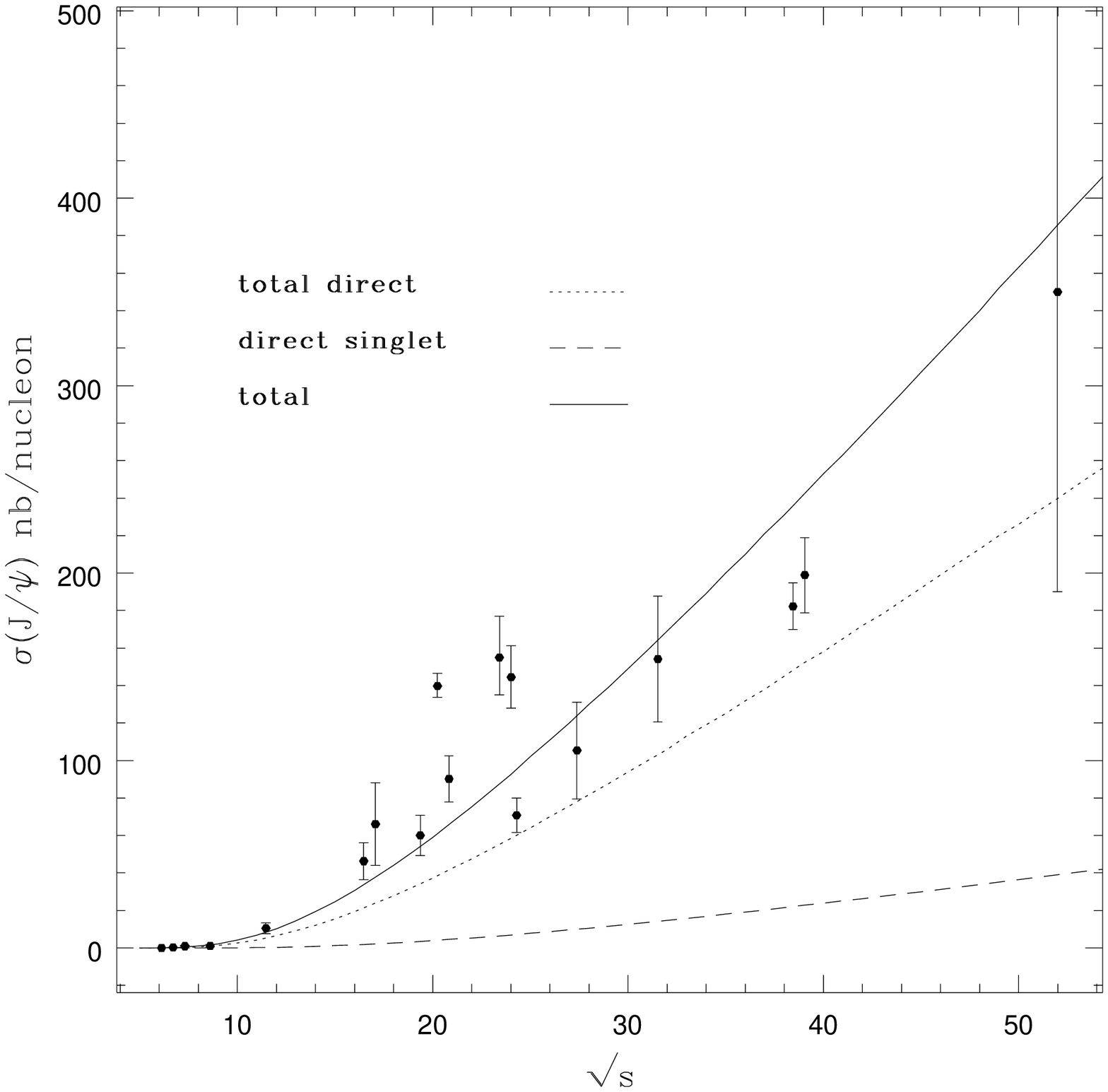,
             height=7cm, width=8cm,
            clip=
}
\end{minipage}
\begin{minipage}{8cm}
\epsfig{file=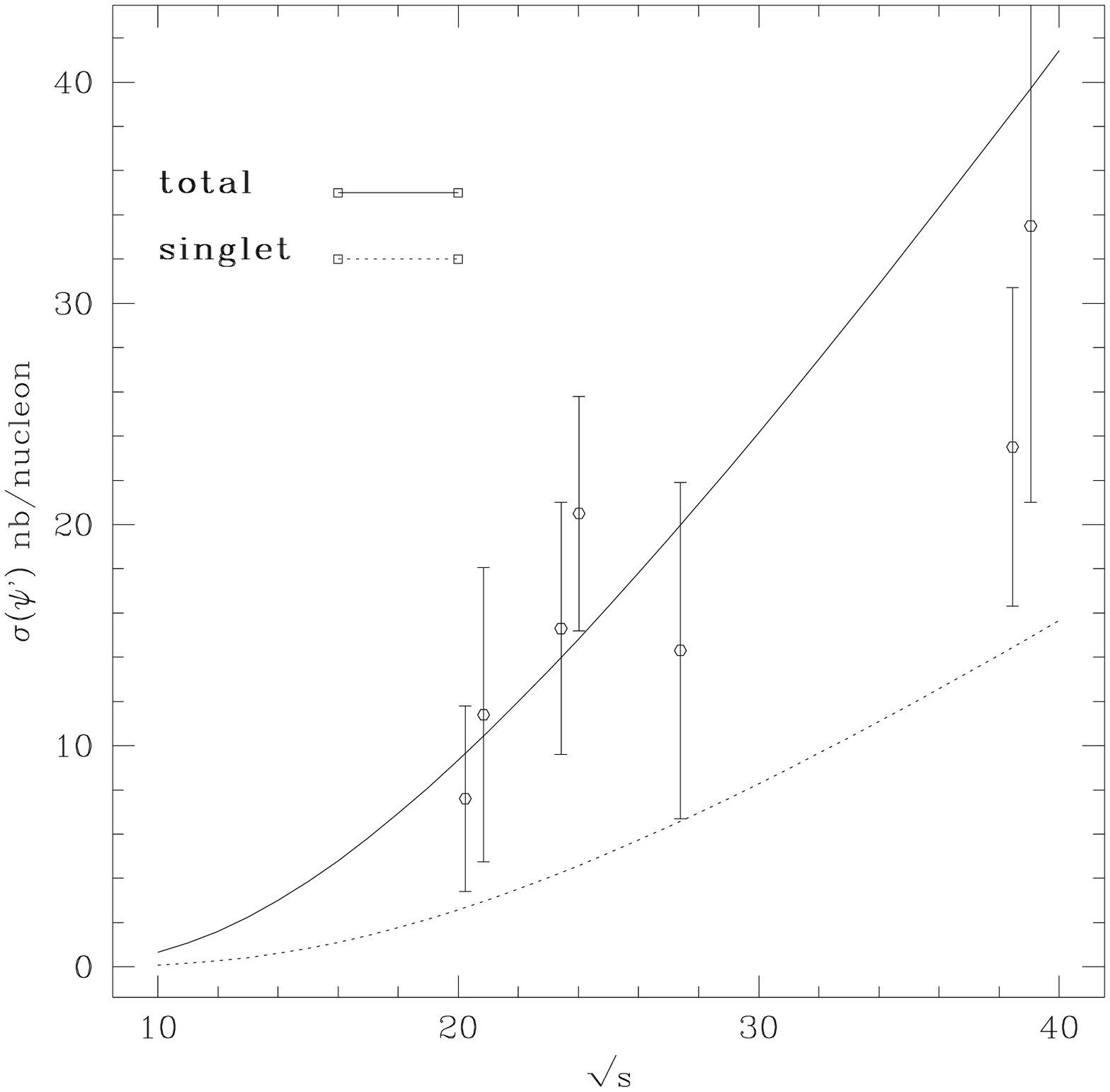,
             height=7cm, width=8cm,
            clip=
}
\end{minipage}
\caption{\label{fig3}
\small $\jpsi$ (left) and $\psi'$ (right) production in $pN$ fixed
target experiments, compared with theoretical fits with the color octet
contributions \protect\cite{br}.}
\end{center}
\end{figure}

Fixed target experiments were of course the first to produce  data
on quarkonia production. Large normalization discrepancies between, say,
$\jpsi$ production data and theoretical predictions based on the Color Singlet
Model had been observed but, lacking a detailed understanding of the problem,
usually dealt with by scaling the theoretical curves by very large K-factors,
of order ten or more. Nowadays, the FA offers the possibility to give a
theoretically sound interpretation of these data, possibly in line with the one
which seems to successfully describe the Tevatron data. Color octet
contributions to total $\jpsi$ and $\psi'$ production in a fixed target 
set-up have been
evaluated, and compared to data obtained in $pN$ and $\pi N$ collisions. 
Fitting $p N$ data with leading order short distance cross sections (a
next-to-leading order calculation is in preparation \cite{cgmmp}) 
the following values for the matrix elements have been found \cite{br}:
\beqa
&&\langle\o^\jpsi_8(\oneSzero)\rangle 
+ {7\over m^2} \langle\o^\jpsi_8(^3P_0)\rangle
\simeq 3 \times 10^{-2}~{\rm GeV}^3 \qquad \mbox{\cite{br}}\\
&&\langle\o^{\psi'}_8(\oneSzero)\rangle 
+ {7\over m^2} \langle\o^{\psi'}_8(^3P_0)\rangle
\simeq 5.2 \times 10^{-3}~{\rm GeV}^3 \qquad \mbox{\cite{br}}
\eeqa
Figure \ref{fig3} shows how such values for the color octet matrix elements
allows for a good description on the data in a wide beam energy range, whereas
the singlet contribution alone clearly underestimates them. A similar analysis
has also been performed in \cite{gs}. It is to be noted
that the fitted values are not fully consistent with the Tevatron ones, looking
at least a factor of three smaller (notice that the linear combination of the
two matrix elements is not exactly the same): this can be taken as a further
indication that the Tevatron fits might be an overestimate).

These parameters fitted in $pN$ collisions have been tested by predicting
$\jpsi$ and $\psi'$ production in $\pi N$ collisions: they provide a fairly
good description of the data, though they seem to underestimate them by about 
a factor of two \cite{br}.

A few more observables can offer a good handle on the validity of a given
quarkonium production framework, namely $\chi$'s production, relative fraction
of $\psi$'s and $\chi$'s, and polarization of the produced quarkonium. Let us
consider them in turn.

The ratio $\sigma(\chi_1)/\sigma(\chi_2)$ has been measured in many
experiments, both in $pN$ and in $\pi N$ collisions. An average
value for $\pi N$ is about 0.6, while two $p N$ experiments give smaller
values ($0.34 \pm 0.16$ for E771, $0.24\pm 0.3$ for E673) but with large
errors. A new preliminary result, $0.45 \pm 0.2$, has been presented by E771 
at this Conference \cite{gollwitzer}.

These experimental results have to be compared with a theoretical prediction,
within the FA, of about 0.08 \cite{br}. The reason for such a small prediction
is that at leading order the $gg\to ^3\!P_1^{(1)}\to\chi_1$ process vanishes in
the FA, and the color octet channel $q\bar q\to\threeSoneo\to\chi_J$ only
contributes a small fraction. One should however consider that more color
octet channels can contribute to both $\chi_1$ and $\chi_2$ production. These
processes go through the octet states $\oneSzeroo$, $\threePjo$ and
$^3D_J^{(8)}$, and an accurate assessment of their relevance is prevented by
our ignorance of the values of the NRQCD matrix elements weighing their
transition to the observable $\chi$ states. A very crude estimate of their
effect \cite{br,beneke} returns a value around 0.3, in good agreement with the
$pN$ data but smaller that the $\pi N$ ones\footnote{Of course this
disagreement between the two sets of data, if found to persist and become more
significant, would by itself be pretty puzzling.}.

One more interesting observable is the fraction of $\jpsi$ coming from
$\chi$'s decays, $\sigma(\jpsi\leftarrow\chi)/\sigma(\jpsi)$, which is of
course also a part of the ratio of $\jpsi$ and $\chi$'s cross sections. The
experimental data, both in fixed target experiments in $pN$ and $\pi N$
collisions and also in $p\bar p$ collisions at the Tevatron, gather around a
central value of 0.3--0.4. The FA, making use of fits to Tevatron data for the
$\chi$'s matrix elements and to fixed target data for $\jpsi$ ones  predicts a
value of about 0.3 \cite{br}, thus in good agreement with the experiment.

What's special about this observable is that its value is strikingly different
in $\gamma p$ collisions. This is because the leading order reaction $\gamma
g\to \chi_J g$ is forbidden due to charge conjugation invariance. Indeed, no
experimental data exist for $\sigma(\jpsi\leftarrow\chi)/\sigma(\jpsi)$ in
photoproduction, due to the vanishingly small $\chi$'s yield, but only an upper
limit by NA14 \cite{na14}: $\sigma(\jpsi\leftarrow\chi)/\sigma(\jpsi)<0.08$.
The reason why this is important is that the Color Evaporation Model, in that
it is not concerned with the details of the particles initiating the reaction,
would predict the same ratio for hadro- and photoproduction. Hence, one could
say that the CEM is ruled out by this result, though more independent
confirmations of the photoproduction experimental result would actually be
welcome.

One further key observable is the polarization of the quarkonium, measured
according to eq. (\ref{pol}). One finds, experimentally,
\beqa
&&\alpha(\jpsi) = 0.02  \pm 0.14\\
&&\alpha(\psi') = 0.028 \pm 0.004\qquad
\eeqa
That is, the quarkonia are found to be produced essentially unpolarized.

This is in contrast with theoretical calculations within the FA, which
predict instead a sizeable degree of polarization, returning
$0.31<\alpha(\jpsi)<0.62$ and $0.15<\alpha(\psi')<0.44$ \cite{br}, the large
band taking into account the very approximate knowledge we have of the NRQCD
matrix elements' values.

Such a discrepancy is certainly disturbing, and if confirmed would be a serious
problem for the factorization approach. At the present stage of our
understanding we must however be aware that many unaccounted for contributions
may still play an important role here. For instance, a proper inclusion of 
effects which would lead to off-shell rather than on-shell colliding gluons
could significantly change the picture presented above. I therefore think it
would not be wise to bury the factorization approach at this stage and because
of this discrepancy.

\section{Conclusions}

With this brief survey of a few of the phenomenological consequences of the
factorization approach to quarkonia production (which I should urge you {\sl
not} to call ``Color Octet Model''!) we have seen how it looks able to describe
in a satisfactory way experimental results previously at great variance with
the Color Singlet Model, both in fixed target experiments and at the Tevatron.

Problems appear in photoproduction at HERA, where the color singlet
contribution alone appears on the other hand to well describe the data. 
But the uncertainties are
still large and can accomodate for the discrepancy. 

Polarization data are also troublesome for the factorization approach (as they 
are for the CSM, while the CEM cannot give a prediction at all). But I feel,
once again, the uncertainties to be still too large for an assessment of the
validity of this approach based on these data.

Surely enough, the factorization approach appears superior to both the CSM 
(of which it is an extension) and the CEM. One could say the CSM approximation
still appears to work fairly well when few gluons are involved (in
photoproduction), whereas the CEM can work when very many gluons are around
(in hadron-hadron collisions). But neither of them can even attempt to
describe the whole yield of data: for instance, the CSM can be ruled out by
the Tevatron data alone, and the CEM could possibly be ruled out by $\chi_J$
photoproduction.

It is quite a widespread belief (though by no means universal!) that the
factorization approach {\sl is} the right theory for heavy quarkonia production
and decay. As a matter of fact, the NRQCD lagrangian on which it is based is
nothing but a limit of the QCD lagrangian itself, and not an {\it ad hoc}
model. 

If anything, a problem of the FA is not of being inadequate, but rather of
being perhaps even too general. Many different NRQCD matrix elements enter the
phenomenological predictions (because the corresponding operators enter the
nonrelativistic limit of the QCD lagrangian), and it is difficult to produce
accurate numerical results with so many unknown parameters. These matrix
elements are on the other hand rigorously defined and in principle calculable
by lattice QCD, so it is possible they will be more precisely determined in the
future either this way or by global  analyses of experimental data.

Given the reasonable correctness of the underlying lagrangian, discrepancies of
the theoretical predictions with experimental data may still be originated by
the approximations included in the calculations.

Higher twist corrections to the factorization approximation can be large,
especially for charmonium (one should never forget that $\Lambda/m_c \simeq
0.3$, not really a negligibly small number). Higher order corrections, both in
the strong coupling ($\alpha_s(m_c) \simeq 0.3$) and in the velocity of the
heavy quarks (again, $v^2 \simeq 0.3$ for charm), can also be large. The value
of $v^2$ (and higher powers) determines -- via scaling rules \cite{LMNMH92} --
which matrix elements are dominant, and also how accurately -- via spin
symmetry -- different matrix elements can be equated to each other. These
approximations are widely used in the phenomenological predictions, to truncate
the series and to decrease the number of independent parameters. How reliable
they are is therefore extremely important for the accuracy of the theoretical
result: the fairly large value of $v^2$ for charmonium systems may help
explaining discrepancies between nowadays predictions and experimental data.

The situation should be much better for bottomonium systems, for which all the
expansion parameters I mentioned take significantly smaller values, around 0.1.
All the approximations should therefore be much better justified, and one
should expect a better agreement between theory and data. A detailed study of
such systems, both theoretically and experimentally, will therefore greatly
help finally confronting the factorization approach to quarkonia production and
decay with the real world.

\vspace{.5cm}
\noindent
{\bf Acknowledgements.} I wish to thank the Organizers of this Conference,
Patrick Aurenche for inviting me to give this talk, and all the participants
for the pleasant atmosphere I enjoyed during my stay.

\end{document}